\documentstyle[12pt,epsf]{article}
\newcommand{\be}{\begin{equation}}
\newcommand{\ee}{\end{equation}}
\newcommand{\ben}{\begin{eqnarray}\displaystyle}
\newcommand{\een}{\end{eqnarray}}

\newcommand{\sectiono}[1]{\section{#1}\setcounter{equation}{0}}
\renewcommand{\theequation}{\thesection.\arabic{equation}}
\def\ket#1{\left| #1\right\rangle}

\catcode`\@=11
\def\marginnote#1{}
\newcount\hour
\newcount\minute
\newtoks\amorpm
\hour=\time\divide\hour by60
\minute=\time{\multiply\hour by60 \global\advance\minute by-\hour}
\edef\standardtime{{\ifnum\hour<12 \global\amorpm={am}%
        \else\global\amorpm={pm}\advance\hour by-12 \fi
        \ifnum\hour=0 \hour=12 \fi
        \number\hour:\ifnum\minute<10 0\fi\number\minute\the\amorpm}}
\edef\militarytime{\number\hour:\ifnum\minute<10 0\fi\number\minute}
\def\draftlabel#1{{\@bsphack\if@filesw {\let\thepage\relax
   \xdef\@gtempa{\write\@auxout{\string
      \newlabel{#1}{{\@currentlabel}{\thepage}}}}}\@gtempa
   \if@nobreak \ifvmode\nobreak\fi\fi\fi\@esphack}
        \gdef\@eqnlabel{#1}}
\def\@eqnlabel{}
\def\@vacuum{}
\def\draftmarginnote#1{\marginpar{\raggedright\scriptsize\tt#1}}
\def\draft{\oddsidemargin -.5truein
        \def\@oddfoot{\sl preliminary draft \hfil
        \rm\thepage\hfil\sl\today\quad\militarytime}
        \let\@evenfoot\@oddfoot \overfullrule 3pt
        \let\label=\draftlabel
        \let\marginnote=\draftmarginnote
   \def\@eqnnum{(\theequation)\rlap{\kern\marginparsep\tt\@eqnlabel}%
\global\let\@eqnlabel\@vacuum}  }

\def\titlepage{\@restonecolfalse\if@twocolumn\@restonecoltrue\onecolumn
     \else \newpage \fi \thispagestyle{empty}\c@page\z@
        \def\thefootnote{\fnsymbol{footnote}} }

\def\endtitlepage{\if@restonecol\twocolumn \else  \fi
        \def\thefootnote{\arabic{footnote}}
        \setcounter{footnote}{0}}  

\catcode`@=12
\relax
\def\half{\frac{1}{2}}
\def\beq{\begin{equation}}
\def\eeq{\end{equation}}

\newcommand{\real} {{{\rm I} \kern -0.2em {\rm R}}}
\newcommand{\complex} {{{\sf I} \kern -0.48em {\rm C}}}
\newcommand{\naturel} {{{\rm I}  \kern -0.18em {\rm N}}}
\newcommand{\integer} {{{\rm Z} \kern -0.31em {\rm  Z}}}
\newcommand{\smallinteger} {{{\rm Z} \kern -0.25em {\rm  Z}}}

\relax

\def\crbig{\\\noalign{\vspace {3mm}}}

\relax

\begin{document}
\setcounter{footnote}{0}
\setcounter{page}{1}

{}~
\hfill\vbox{\hbox{hep-th/0008101}\hbox{CTP-MIT-3013}
}\break

\vskip 2.0cm

\centerline{\large \bf Codimension two lump solutions in string field theory}
\centerline{\large \bf and tachyonic theories}

\vspace*{6.0ex}

\centerline{\large \rm Nicolas Moeller}

\vspace*{6.5ex}

\centerline{\large \it Center for Theoretical Physics}

\centerline{\large \it
Massachusetts Institute of Technology}

\centerline{\large \it Cambridge,
MA 02139, USA}
\vspace*{1ex}
\centerline{E-mail: moeller@pierre.mit.edu}

\vspace*{4.5ex}

\centerline {\bf Abstract}
\bigskip
We present some solutions for lumps in two dimensions in level-expanded string 
field theory, as well as in two tachyonic theories: pure tachyonic string field 
theory and pure $\phi^3$ theory. Much easier to handle, these theories might 
be used to help understanding solitonic features of string field theory. We 
compare lump solutions between these theories and we discuss some 
convergence issues.
\vfill \eject

\baselineskip=18pt

\sectiono{Introduction} \label{s0}

In the last few months, there has been growing evidence that 
level truncation is a good way of doing computations in string field
theory. In particular, it allows to get very accurate results for the
string field theory true vacuum, both in open bosonic string field theory
and superstring field theory (\cite{MSZ} - \cite{0006240}).

More recently, in \cite{MSZ}, a level truncation scheme has been developed
which takes non-zero momentum into account; Applied to lump solutions in 
one
dimension, it gave numerical results for the ratio of the tension of a
D-$p$-brane and a D-$(p-1)$-brane with a precision of about 1\%. In \cite{MELLO},
de Mello Koch and Rodrigues applied this scheme to construct 2-dimensional 
lumps in open bosonic string field theory.

In this paper, we want to present
independent results on 2-dimensional lumps. We also describe these lumps in
two theories involving only the tachyon: pure tachyonic sft 
(string field theory in which we keep only the tachyon, including its higher 
derivatives), with action (\cite{0003031}): 
\beq
S = - 2 \pi^2 T_{25} \int d^{26} x \, \left( \half \partial_{\mu} \phi \, 
\partial^{\mu} \phi - \half \phi^2 + {1 \over 3} K^3 \tilde{\phi}^3 \right)\ ,
\eeq
where $T_{25}$ is the D-$25$-brane tension, 
$K = 3 \sqrt{3} / 4$ and $\tilde{\phi} = K^{\partial_{\mu} \partial^{\mu}} 
\phi$.
And pure $\phi^3$ theory 
(the usual scalar $\phi^3$ theory of a tachyon, which doesn't include higher 
derivatives), with action
\beq
S = - 2 \pi^2 T_{25} \int d^{26} x \, \left( \half \partial_{\mu} \phi \,
\partial^{\mu} \phi - \half \phi^2 + {1 \over 3} K^3 \phi^3 \right)\ ,
\eeq
the only difference being that here we have $\phi^3$ instead of $\tilde{\phi}^3$.

\sectiono{Calculating the potential} \label{s1}

We will use the notation of \cite{MSZ}, but we will compactify two dimensions, 
instead of one, on a torus. Let us name $x$ and $y$ the compact dimensions. We 
impose the identifications
\beq
\begin{array}{rcl}
x & \sim & x + 2 \pi R \ , \cr
y & \sim & y + 2 \pi R \ .
\end{array}
\eeq
The $x$- and $y$-momenta will be quantized:
\beq
\begin{array}{rcl}
p_x &=& m/R \ , \cr
p_y &=& n/R \ .
\end{array}
\eeq
For each zero-momentum state $\ket{\Phi_i}$ that appears in the non-compact theory, 
we will have states labeled by two indices $\ket{\Phi_{i, mn}}$ with levels 
\beq
l(\Phi_{i, mn}) = l(\Phi_i) + (m^2 + n^2)/R^2 \ .
\eeq
By definition, when we work at level $(M, N)$, we keep fields of level $\leq M$, 
and terms in the potential of total level $\leq N$. In this paper, we will work 
in string field theory at level (2, 4) and in pure tachyonic theories at arbitrary 
levels. Therefore all the fields we need are:

\beq
\begin{array}{rcl}
\ket{T_{mn}} & = & 
\displaystyle{
\left\{
 \begin{array}{l}
c_1 \, \cos{\left({m \ x \over R}\right)} \ \cos{\left({n \ y \over R}\right)} \ket{0} 
\  , \ \ m=n \cr
c_1 \, \left( \cos{\left({m \ x \over R}\right)} \ \cos{\left({n \ y \over 
R}\right)} +
\cos{\left({n \ x \over R}\right)} \ \cos{\left({m \ y \over R}\right)} \right) 
\ket{0} \  , \ \ m \neq n
\end{array}
\right.
} \crbig
\ket{U_{00}} & = & c_{-1} \ket{0} \crbig
\ket{V_{00}} & = & c_1 \left( L_{-2}^X + L_{-2}^Y \right) \ket{0} \crbig
\ket{W_{00}} & = & c_1 L_{-2}^{\prime} \ket{0},
\end{array}
\eeq
where $L_{-2}^X$ and $L_{-2}^Y$ are Virasoro generators of the CFT of the 
compact dimensions $x$ and $y$ respectively, and $L_{-2}^{\prime}$ is a Virasoro 
generator of the CFT of the 24-dimensional co-space. 
The definition of $\ket{T_{mn}}$ ensures that the solutions we will find 
are symmetric under $x \leftrightarrow y$ as well as under  
$x \rightarrow -x$ and  $y \rightarrow -y$. 
We will thus use the string field:

\beq
\ket{\vec{T}} = \sum_{m \leq n} t_{mn} \ket{T_{mn}} + u_{00} \ket{U_{00}} + 
v_{00} \ket{V_{00}} + w_{00} \ket{W_{00}}, 
\eeq
where the sum is restricted by the level truncation, the level of each field being

\beq
l(T_{mn}) = (m^2 + n^2) / R^2 \ \ , \ \ l(U_{00}) = l(V_{00}) = l(W_{00}) = 2 \ .
\eeq

We will not repeat here how to calculate the potential ${\cal V}_{MN} 
(\vec{T})$ at level $(M, N)$. We refer to the literature (see for example 
\cite{MSZ}, 
\cite{9912249}, \cite{0001201}, \cite{0002237}, \cite{0006240}). Note however that 
in theories involving only the tachyon, the coefficients of the terms in the 
potential can be
calculated straightforwardly, the only difficulties being to keep track of 
momentum conservation at the vertex and to figure out the combinatorial 
factors.
We have written computer codes calculating the potentials in the following 
theories:

\begin{itemize}
\item string field theory at arbitrary radius up to level (2, 4).
\item pure tachyonic sft at arbitrary radius and arbitrary level.
\item pure $\phi^3$ theory at arbitrary radius and arbitrary level.
\end{itemize}

\sectiono{Codimension 2 lumps} \label{s2}

\subsection{String field theory truncated at level (2,4)}

Before showing results, let us say a few words about how to find these lumps given 
the potential. In general, the tachyon potential has many extrema. In doing
computations at high level, it might be difficult to setup the convergence
on the right branch. The method we've used here \footnote{I thank B. Zwiebach 
for this suggestion.} is to go backwards: We know
the approximate shape of the lump we are looking for, from it one can
calculate its Fourier expansion. We can then plug these Fourier coefficients
into our algorithm finding solutions of the
equations of motion (by Newton's method). If we start the
numerical algorithm with a seed close enough to the solution we want,
it is fairly probable that it will converge to the sought solution.

\begin{figure}[!ht] 
\leavevmode
\begin{center}
\epsfbox{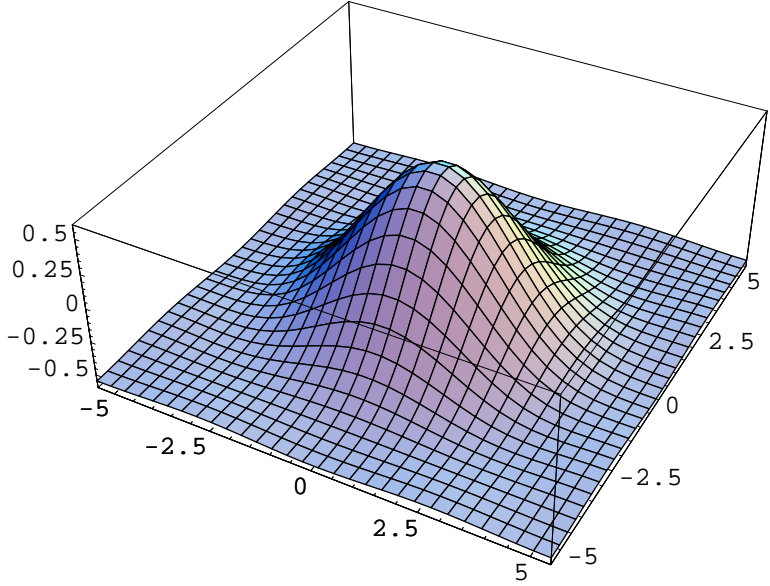}
\end{center}
\caption{A lump in sft at level (2, 4) with $R=\sqrt{3}$. The plot represents $-t(x, y)$ 
as a function of $x$ and $y$.} \label{sftRsqrt3}
\end{figure}

Let us present our result in string field theory at level $(2,4)$.
In order to compare our result with \cite{MELLO}, we have computed it at
$R=\sqrt{3}$. Fig.\ref{sftRsqrt3} shows a plot of the solution.

\noindent The ratio of the tension of a D-$p$-brane and a D-$(p-2)$-brane, 
when divided by its
expected value $(2 \pi)^2$, can be approximated by (\cite{MSZ}):
\beq
r^{(2)} \equiv R^2 \, ( 2 \pi^2 {\cal V}_{(M, N)} 
( \vec{T}_{lump}) - 2 \pi^2 {\cal V}_{(M, N)}
(\vec{T}_{vac})),
\eeq
where $\vec{T}_{lump}$ is the lump solution, $\vec{T}_{vac}$ is the solution 
corresponding to the true vacuum at the given truncation level $(M, N)$ and 
${\cal V}_{(M, N)}$ is the potential at level $(M, N)$. With our solution above, 
we find
\beq
r^{(2)}_{(2, 4)} = 1.13025 \ ,
\eeq
13\% away from the expected value of unity.  \footnote{In \cite{MELLO}, 
the authors find 1.1378 at
the same truncation level. There is a slight disagreement between these two 
numbers, though it might be due to round off or numerical error.}

\begin{figure}[!ht] 
\leavevmode
\begin{center}
\epsfbox{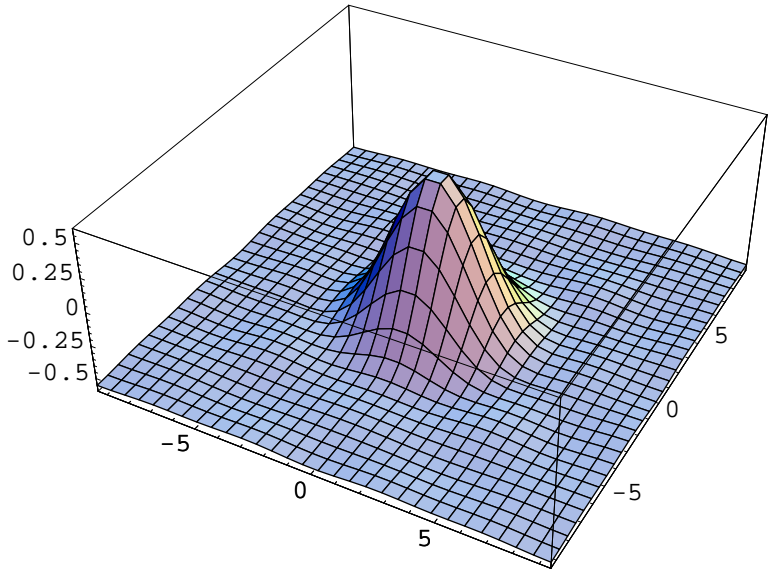}
\end{center}
\caption{A lump in sft at level (2, 4) with $R=3$. The plot represents $-t(x, y)$ 
as a function of $x$ and $y$.} \label{sftR3}
\end{figure}

To show that the size of the lump is, in the large radius limit, independent of 
the radius, let us 
compare with the solution at $R=3$ (Fig.\ref{sftR3}).

\begin{figure}[!ht] 
\leavevmode
\begin{center}
\epsfbox{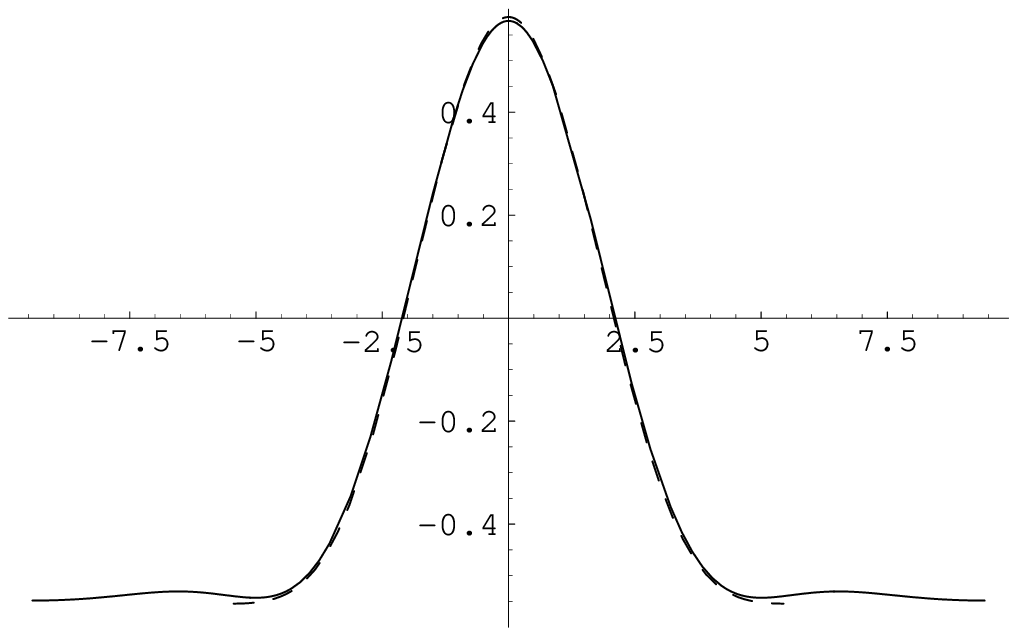}
\end{center}
\caption{The dashed curve is the profile $-t(x,0)$ of the lump solution in sft 
at $R=\sqrt{3}$, 
the solid curve is the profile $-t(x,0)$ of the lump solution at $R=3$} \label{comp}
\end{figure}

\noindent To compare the sizes of this lumps, we plot together their profiles 
$-t(x,0)$. Fig.\ref{comp} clearly shows the radius-independence of the 
shape of the lump.

\subsection{Pure tachyonic string field theory and pure $\phi^3$ theory}

\begin{figure}[!ht] 
\leavevmode
\begin{center}
\epsfbox{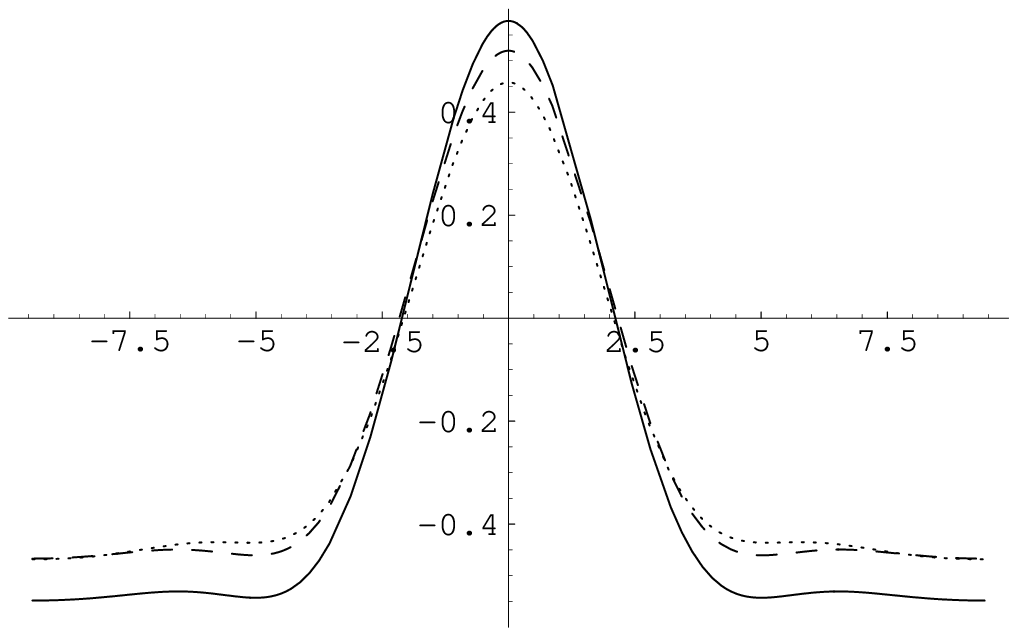}
\end{center}
\caption{Lump profiles $-t(x,0)$ in string field theory (solid curve), 
pure tachyonic sft (dashed curve) and 
pure $\phi^3$ theory (dotted curve) at level (2, 4) and $R=3$.} \label{threecomp}
\end{figure}

We do find codimension 2 lump solutions in these theories. In fig.\ref{threecomp}, 
we show the profiles $-t(x,0)$ of the lumps in the three different theories. 
We have taken the three lumps to be at level (2, 4) with $R=3$. The different 
asymptotic values of $t(x, 0)$ show the different vev's of the tachyon in the 
three theories.

\begin{figure}[!ht] 
\leavevmode
\begin{center}
\epsfbox{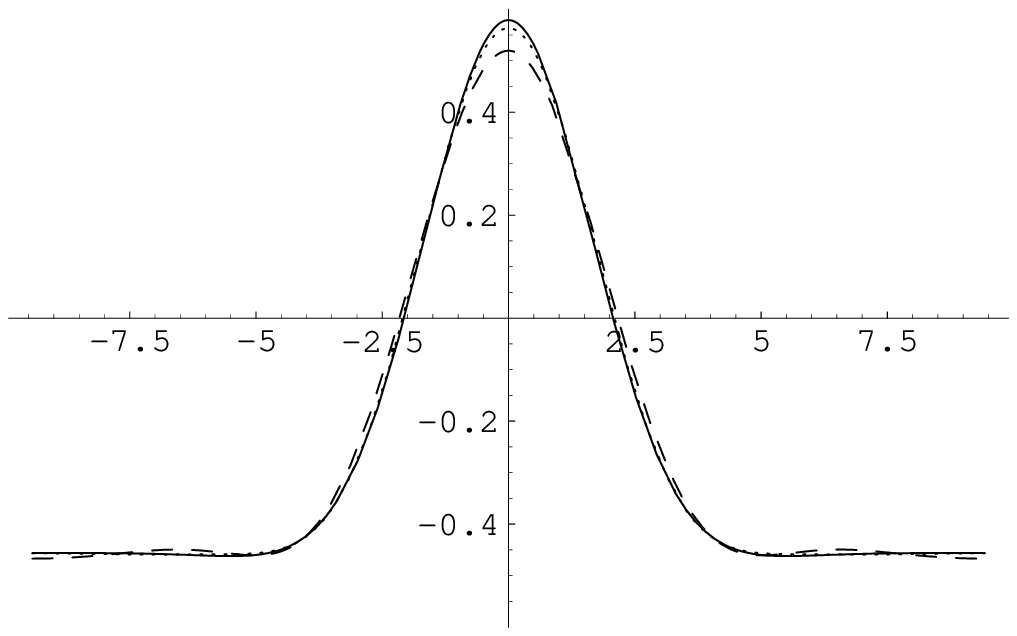}
\end{center}
\caption{Lump profiles $-t(x,0)$ in pure tachyonic sft with $R=3$, at level 
(2, 4) (dashed curve), (3, 9) (dotted curve) and (10, 30) (solid curve)} \label{levtach}
\end{figure}

\begin{figure}[!ht] 
\leavevmode
\begin{center}
\epsfbox{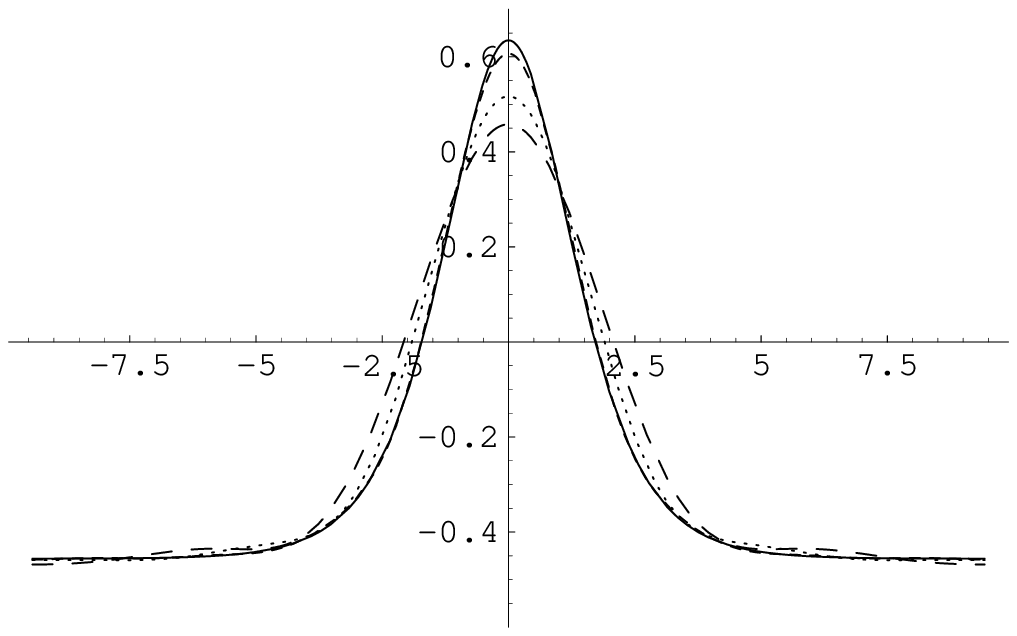}
\end{center}
\caption{Lump profiles $-t(x,0)$ in pure $\phi^3$ theory with $R=3$, at level 
(2, 4) (big-dashed curve), (3, 9) (dotted curve), (7, 21) (small-dashed curve), 
and (20, 60) (solid curve)} \label{levphi3}
\end{figure}

The pure tachyonic theories are much more tractable for numerical computations. 
We have written codes giving the actions at arbitrary level. To illustrate the 
convergence of the level truncation scheme, we show in figs.\ref{levtach},
\ref{levphi3} 
the lump profiles at different truncation levels.

It is interesting to see that the solution in pure $\phi^3$ theory converges much 
slower than in pure tachyonic sft. This is due to the fact that in pure 
tachyonic sft, the coefficient of the term $t_{m_1 n_1} t_{m_2 n_2} 
t_{m_3 n_3}$ in the potential is proportional to 
$K^{3 - (m_1^2 + n_1^2 + m_2^2 + n_2^2 + 
m_3^2 + n_3^2) / R^2}$. Since $K>1$, at high level these terms are much less 
important than 
in pure $\phi^3$ theory where the same coefficients are proportional to $K^3$.

\sectiono{Conclusion}

We do find codimension 2 lumps solutions in all three theories considered in 
this paper. Note that there is an apparent conflict with Derrick's theorem which 
states that solitons in scalar field theory can exist only in codimension $<2$. 
But one 
of the assumptions used in the proof of the theorem is that the potential must 
be bounded below, which is not the case in the theories considered here 
\footnote{I wish to thank B. Zwiebach for pointing this out.}. 
This negativity allows the existence of solutions, as shown in \cite{0002117}.

As it is easy to use, pure tachyonic sft is an interesting toy model of full 
string field theory. We saw that it converges very fast when the 
truncation level is increased. Moreover we found a lump solution of 
approximately the same shape as in sft, this may show that we can study other 
kinds of sft solitons (like intersecting branes) by using the pure 
tachyonic approximation (work in progress \cite{progress}).

Finally, we saw the amusing fact that string field theory, due to the coefficients 
$K^{3 - {\rm level}}$ in front of every term, seems to be better 
fit for level truncation  
than the simple $\phi^3$ theory which converges very slowly as the 
level is increased.

\vspace{14pt}

\noindent {\bf Acknowledgements:} I would like to thank A. Sen and 
B. Zwiebach for detailed discussions and comparisons, and W. Taylor for discussions. 
This work was supported in part by DOE contract 
\#DE-FC02-94ER40818.


\end{document}